\documentclass[conference]{IEEEtran}
\usepackage[a4paper,left=3cm,right=2cm,top=2.5cm,bottom=2.5cm]{geometry}
\usepackage{cite}
\usepackage[pdftex]{graphicx}
\usepackage[cmex10]{amsmath}
\usepackage{amssymb}

\usepackage{times}
\usepackage{array}
\usepackage[tight,footnotesize]{subfigure}
\usepackage[font=footnotesize]{subfig}
\usepackage{graphicx}
\usepackage{mathtools}
\usepackage{amsmath}

\usepackage{alltt}

\def\boxx{{\vcenter{\vbox{\hrule height.3pt
          \hbox{\vrule width.3pt height6pt
          \kern6pt\vrule width.3pt}\hrule height.3pt}}\;}}

\def\impos{{\;\vcenter{\hbox{\rule{5mm}{0.2mm}}} \vcenter{\hbox{\rule{1.5mm}{1.5mm}}} \;}}

\def\lrarrow{\leftrightarrow \kern-8pt \rightarrow}

\def\2{\frac{1}{2}}

\def\beq{\begin{eqnarray}}
\def\eeq{\end{eqnarray}}
\def\2{\frac{1}{2}}

\newtheorem{principle}{Principle}

\def\lrarrow{\leftrightarrow \kern-8pt \rightarrow}

\def\frightarrow{\rightarrow \kern-11pt /~~}
\def\reducesto{\simeq \kern -3pt >}

\usepackage{fancyhdr}					
\begin{document}
\newcommand{\strust}[1]{\stackrel{\tau:#1}{\longrightarrow}}
\newcommand{\trust}[1]{\stackrel{#1}{{\rm\bf ~Trusts~}}}
\newcommand{\promise}[1]{\xrightarrow{#1}}
\newcommand{\revpromise}[1]{\xleftarrow{#1} }
\newcommand{\assoc}[1]{{\xrightharpoondown{#1}} }
\newcommand{\rassoc}[1]{{\xleftharpoondown{#1}} }
\newcommand{\imposition}[1]{\stackrel{#1}{\impos}}
\newcommand{\scopepromise}[2]{\xrightarrow[#2]{#1}}
\newcommand{\handshake}[1]{\xleftrightarrow{#1} \kern-8pt \xrightarrow{} }
\newcommand{\cpromise}[1]{\stackrel{#1}{\frightarrow}}
\newcommand{\policy}{\stackrel{P}{\equiv}}
\newcommand{\field}[1]{\mathbf{#1}}
\newcommand{\bundle}[1]{\stackrel{#1}{\Longrightarrow}}

\title{Koalja: from Data Plumbing to Smart Workspaces in the Extended Cloud}

\author{Mark Burgess and Ewout Prangsma\\~\\Aljabr Inc\\~}
\maketitle
\IEEEpeerreviewmaketitle

\renewcommand{\arraystretch}{1.4}

\begin{abstract}
  Koalja describes a generalized data wiring or `pipeline' platform, built on
  top of Kubernetes, for plugin user code.  Koalja makes the
  Kubernetes underlay transparent to users (for a `serverless'
  experience), and offers a breadboarding experience for development
  of data sharing circuitry, to commoditize its gradual promotion to a
  production system, with a minimum of infrastructure knowledge.
  Enterprise grade metadata are captured as data payloads flow through
  the circuitry, allowing full tracing of provenance and forensic
  reconstruction of transactional processes, down to the versions of
  software that led to each outcome.  Koalja attends to optimizations
  for avoiding unwanted processing and transportation of data, that
  are rapidly becoming sustainability imperatives. Thus one can
  minimize energy expenditure and waste, and design with scaling in
  mind, especially with regard to edge computing, to accommodate an
  Internet of Things, Network Function Virtualization, and more.
\end{abstract}



\section{Introduction} 

In this paper, we propose a revision of the concept of data pipelines
for modernizing and commoditizing the availability of data processing.
There have been many approaches to data processing, from punch cards
to custom chips.  In the days of early monolithic computers, we
thought of data processing as a simple flowchart execution---a serial
approximation to reasoning based on simple machine instructions.
Decades of rethinking processing, driven by the limitations of memory
and CPU led to versions of parallelism, including the use of
client-server and Map-Reduce methods---and the technological ground
beneath the computation has shifted constantly as the economics of
resources evolved.  

Data pipelines emerged as the idea of software
production lines and delivery pipelines became the favoured metaphor,
and attention shifted back to a serial feed-forward model of Distributed
Acyclic Graphs (DAG).  The pure DAG delivery model is quite convenient
from an infrastructure point of view, as it makes the separation of
computation from connectivity simple, but it is correspondingly
simplistic.  Modern processing requires loops and feedback
to perform non-trivial tasks, and feedback implies that flow-control be
{\em data aware}. 

Directed Cyclic Graphs (DCG), i.e.  flowcharts or
Petri Nets\cite{david1} are back in vogue, because they are suitable
for general data processing.  Moreover, the more interactive
client-service model remains an integral and important approach for
farming out function calls (subroutine) to query databases.  This
pushes complexity back onto the user---and the standard approach has
been to assume ever greater programming skills of end-users, to the
point where we've almost shifted to writing tools for an assumed army
of developers.

We shall not give a complete history or exhaustive list of approaches
to data processing here, as our goal is to find a path to `universality' or
commoditization.  The tool space for data processing is vast, and
variegated, from simple tools like `cron' and `make' to simple-minded
tools like Airflow that treat processing as a series of scheduled
tasks without being `data aware'. Some recent references can be found
here \cite{dataflow,wampler1,Li1,Liu1,tachyon,zaharia1}.  

Our work builds on the observations of Maymounkov \cite{petar}, and
his prototype Koji and the Ko language. His work was a reaction to the
ad hoc approaches to managing jobs, with tools insufficient to the
task, at Google. Although an improvement on existing tools, Koji was
still basically developer-oriented, not user friendly without
expertise, and thus we have sought to extend the concepts to absorb
more of the infrastructure tasks and data-awareness into the platform
itself.  We can make a generic and somewhat sweeping generalization to
claim that most cloud technology today has been designed to engage
programmers (developers) and make it easy for them, on a completely
generic API level, rather than make it accessible for end users.  This
leads to a proliferation of vertically optimized third party tools for
enabling very specific tasks in ad hoc ways; these are forced to
implement solutions to a lot of common issues. Our approach is target
those key issues and design a layer on top of Kubernetes (which was
designed principally for stateless services) so that these third party
tools can build on top of an extended set of services, and even
monetize more easily by customization of data wiring and reporting.

\section{Commoditization}

Our goal is to render the experience of collecting and processing data
extremely simple---seeking to balance generality (where no one is
happy with a platform) against specialization (in which services are
insufficient for all), with a compromise in favour of flexible
simplicity over functionality.  Thus, we describe an approach to
building data as generalized circuitry, analogous to electrical
circuitry, in which components are plugin containers for user code,
and wires are scalable data pipelines. 

We start with an approach to describing the wiring principles, with
recognizable data pipelines; then we describe how to turn pipes into
generalized plumbing that can disappear behind the infrastructure wall
of an organization.  Finally we discuss how this smart data plumbing
can be added to a collaborative environment and turned into a
commodity service that we call Workspaces\cite{workspaces}, that integrates
what we currently refer to as Cloud and Internet of Things as a seamless
Extended Cloud.

Amongst the challenges of moving data processing into the cloud, there is
the issue of scaling from large to small, and the relative cost ratio
between cost of processing and cost of infrastructure. The situation of
source data and processing tasks can span a wide area: current cloud
providers are not always even in the same country as users, so data
may have to be transported all the way from an expanding user surface,
at `the edge' of the cloud, to heavily centralized systems with
heavyweight resources in datacentres, and perhaps even back again.
This is a recipe for unnecessary latency and cost---wasteful and
unsustainable. Finally, improvements in the software development
model, involving the containerization of software, also feed into a
remodelling of infrastructure design.  These factors drive us to
reconsider the problem with new constraints.

The question then is: how can we make Wide Area Data Availability and Processing
more accessible to users, without requiring an army of programmers to
couple together fractious tool sets in ad hoc ways in order to support
even the simplest operational requirements? 
We propose a straightforward renormalization of the layers that
underpin data processing to reposition the platforms from being
developer oriented to being user oriented. We dare to claim that the
model can survive shifts in technology, but that might be optimistic.
We have chosen to propose a proof of concept based on the popular
Kubernetes platform layer. Kubernetes is not really a complete platform
for anything, but it serves as a useful {\em lingua franca} for cloud deployments,
across different operating systems and provides.

\section{Koalja}

Our implementation, called Koalja\footnote{It was tempting to call it
  Kojira, in the spirit of renaming everything Go* into Ko*, but the
  obvious translation in Kodzilla sounded fishy.}, realizes a number of
services on top of Kubernetes that extend its model of stateless service pods
into stateful data circuitry. Kubernetes implements a layer of
resource management that is essential to any virtual platform for data
circuitry. It would be meaningless to implement those services,
especially given the popularity of Kubernetes.

\subsection{Design promises}

Two goals stand out in the design:
\begin{itemize}
\item Make the entire cloud become transparent.
\item Add enterprise-grade process observability and traceability.
\end{itemize}
We refer to the model of observational measurement discussed in
\cite{spacetime3,cognitive}, that harks back to
policy compliance and anomaly detection methods pioneered by
CFEngine\cite{burgessC1}, and has since been extended for tracing in
fully distributed systems on a variety of timescales.

If we scratch just below the surface, the key user cases for data
processing are surprisingly straightforward:
\begin{itemize}
\item Data replication and distribution (trivial tasks).
\item Aggregating data (summaries) from multiple sources, often in different regions,
respecting legal and political boundaries, triggered by arrival notifications.
\item Handling aggregation where data come in at different rates from different
sources.
\item Calculating matrix operations for rendering graphics, spreadsheets, and so on.
\item Continuous delivery pipelines (such as software builds), triggered by policy.
\end{itemize}
There are several issues to chew on here, but the key differences lie
in the different scales for data size and inter-arrival time.  It will
initially be difficult to shift industry users out of a `brand'
mentality (one tool for one concern), e.g. Jenkins is The Tool for
software builds, and Kafka is The Tool for data aggregation. Indeed,
each of these tools has characteristics that are adapted to special
tasks. Our aim is to build a skeleton frame into which one could graft
these major organs, or replace them with custom solutions, with a
minimum of programming effort (though, clearly the more one tries to
shoehorn in a special solution, the more workarounds will be
required).  Adaptation to different user cases becomes a matter for policy
and automated adaptation.
The key factors that choose policy are the timescales of the processes. 
\begin{itemize}
\item How fast are observable data changing? 
\item What is the distribution of inter-arrival times (what sampling rate)?
\item How fast a response is needed?
\item How much buffering and collection of data before processing?
\end{itemize}
These are straightforward elements of queueing and---given the freedoms
of cloud services---can be adapted
to without building multiple software projects.

It will take some time to move people away from their traditional
thinking patterns---on the other hand, we are in the midst of this
revolution of thinking, from virtual machines, to containers, to
serverless models, and so on. Some rationalization will be welcome, as
the explosion of APIs and snowballing of complexity associated with it
becomes a significant cost.  For example, industry norms currently try
to distinguish between streaming and batch processing. We do make any
such distinction, but rather provide policies for managing appropriate
timescales.

\subsection{Trigger modes}

The high level promises that support this basic set are:
\begin{itemize}
\item Platform transparency: no reference should ever be made to Kubernetes 
or any other platform in the description of processes. Users should only need
to know their own code, with some rules about how data get dropped off between
tasks.

\item Transparency of operation: users should be able to trace the
  passage and outcomes on an intentional level, like any other
  debugging platform, without having to tap into low level wire
  traffic using proxies like Istio.

\item Tasks should be freely locatable in any region, with transparent interconnection 
between Kubernetes deployments.
\end{itemize}
These promises sound straightforward, but they go against the design principles
of Kubernetes in a number of ways. Kubernetes was not designed with an extended
cloud in mind; nonetheless, it is probably the best current tool for the job. 

\begin{figure}[ht]
\begin{center}
\includegraphics[width=7.5cm]{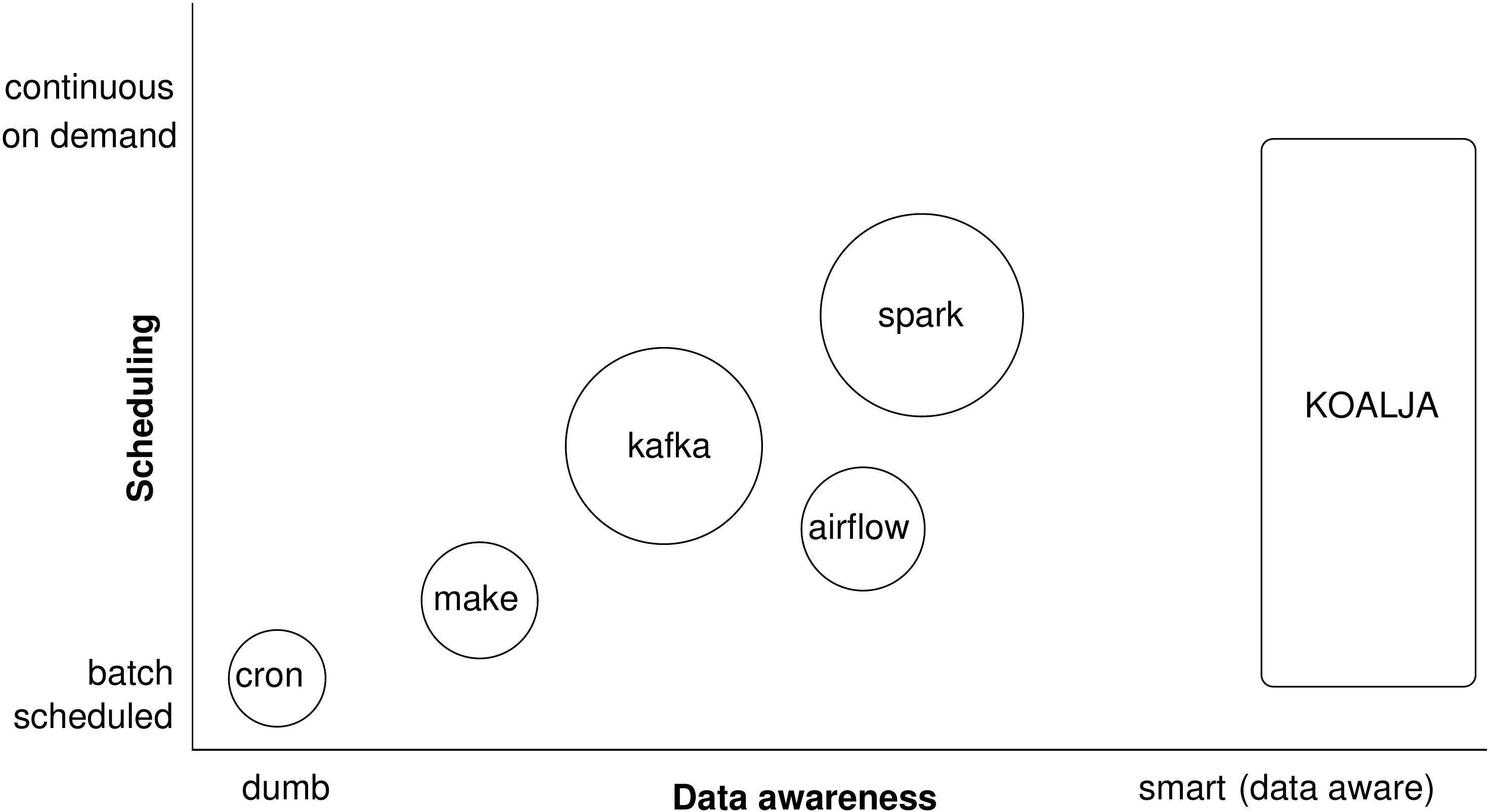}
\caption{\small A skeletal classification of data pipe processing.\label{pipetech}}
\end{center}
\end{figure}

The aim is also to support processing which is triggered by conditions on either end of the pipeline:
\begin{itemize}
\item A `make' model, in which a request for the target at the logical output end 
of the pipes triggers a hierarchical rebuild of dependencies `backwards', recursively.

\item A `reactive' model, in which events arriving at the logical input
end of the pipes triggers when data are pushed into the source
end of the circuitry (driving computation downstream). 
\end{itemize}
These two cases sound orthogonal but they are not, as long as the
causal messaging channel is independent of the data flow itself. In
all cases, whether superficially push or pull on a large scale, tasks
poll or sample some kind of sensor for changes.  Currently, it is
usual to build different systems to handle these cases (figure
\ref{pipetech}), but we feel this situation can be rationalized to
reduce and manage complexity.

The basic architectural elements of a Koalja deployment are thus:
\begin{itemize}
\item Tasks, where users plug in their code for processing.
\item Links, that connect tasks and provide notifications.
\item Storage where actual data batches can be kept and cached for possible re-use,
in all intermediate stages (as with a Makefile process).
\item A pipeline manager that handles registration of processes, scheduling of
work and assembly of metadata.
\end{itemize}

\subsection{Traceability}\label{meta}

\begin{figure}[ht]
\begin{center}
\includegraphics[width=6cm]{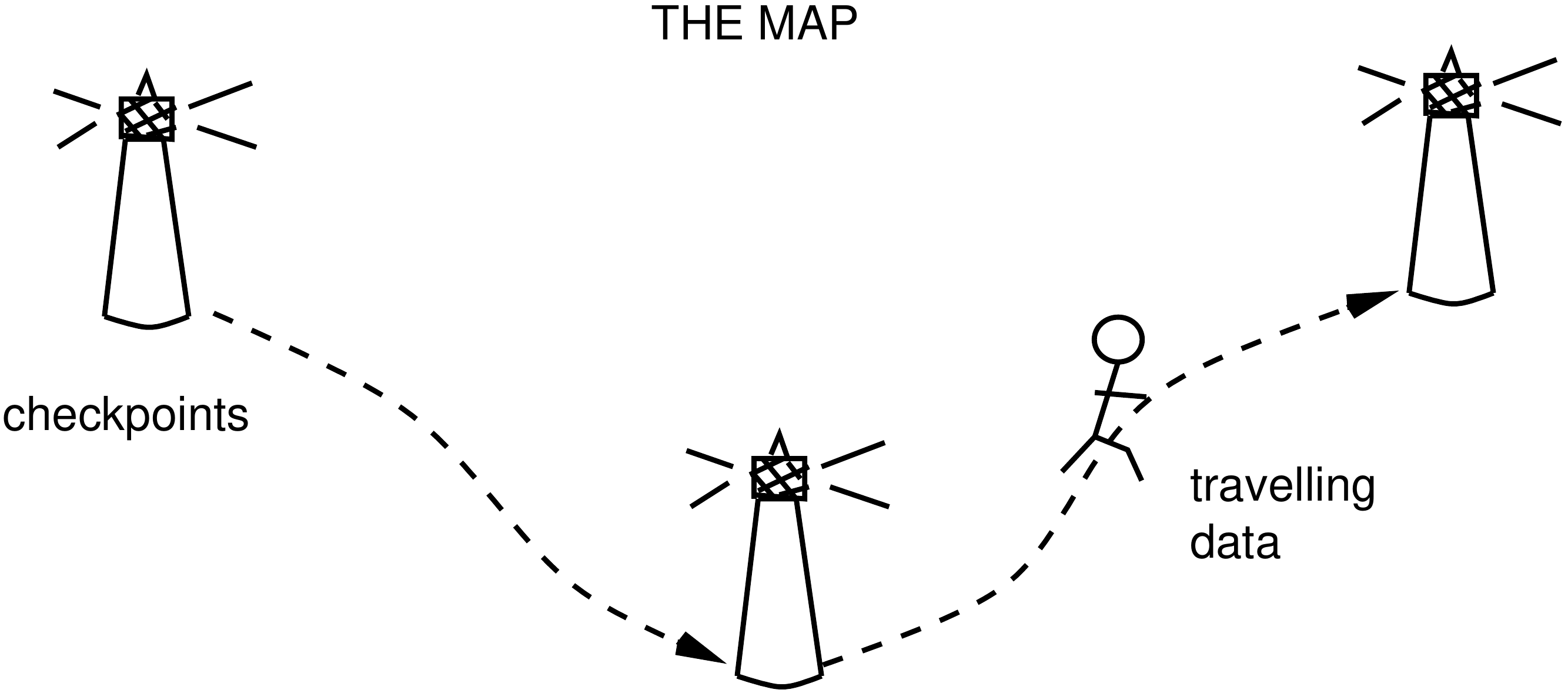}
\caption{\small 3 Views. Travelling passport documents, versus logs of entry and exit from a checkpoint, versus
the map of checkpoints and routes.\label{views}}
\end{center}
\end{figure}

On the tracing side, there are three kinds of story we want to be able to tell about
data processing (figure \ref{views}):
\begin{enumerate}
\item The data traveller log of each transacted data packet or artifact, i.e.
  What a travelling data packet experiences along its journey, which
  software version processed it and in what order?  This is vital to
  diagnosing design flaws, errors, and faults in the outcomes, and for
tracing issues back to their sources.

\item The checkpoint visitor log, as experienced by the user code;
  which data packets and events passed through the checkpoint, and when.
What was done to them?

\item Finally the long term design map that explains the intended
  relationships between the component elements and concepts involved
  in the business process. This includes the topology of checkpoints
  and what promises they make, the kinds of data passed between them,
  significant anomalies, and so forth. We include both the semantics
  of the data, software, and invariant qualities within the system's
  horizon.
\end{enumerate}

\subsection{Melding pipelines with an exterior service model}

\begin{figure*}[ht]
\begin{center}
\includegraphics[width=12cm]{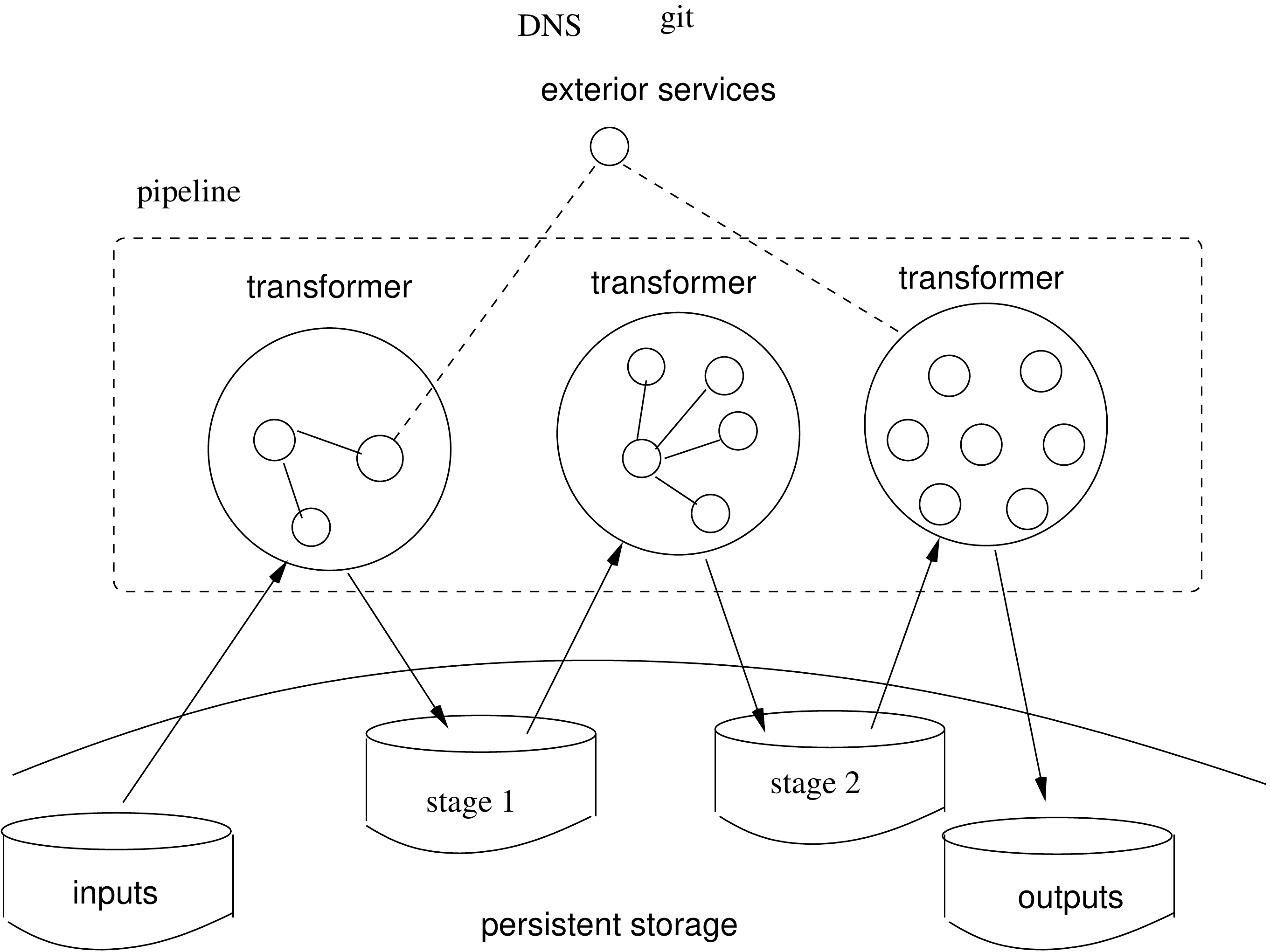}
\caption{\small A pipeline is formed from `data wiring' that forms
sequences passing data between transformations supplied by user code.
These transformations may rely on exterior services, and the certainly rely
on external storage.\label{pipeline}}
\end{center}
\end{figure*}

The elephant in the cloud regarding data pipelines is that a classical
pipeline model all but ignores the most obvious network paradigm of
all: the client-server model. The bulk of interactions on the Internet
use this basic paradigm for communicating. Indeed, the tasks in a
pipeline, each hosted on different containers, uses this paradigm to
pass data and messages between them (see figure \ref{pipeline}).

The main difference between a pipeline abstraction and a client-server
abstraction is that, in a pipeline, every service becomes a client of
the next service in line; i.e.  results are labelled `pass it along'
rather than `return to sender'\footnote{The problem of end to end delivery can
be broken down into local client-server interactions (see
\cite{promisebook}).}. Such is the separation of concerns in a modern systems
(even before the advent of microservices) that client-server interactions for
address lookups, database queries, and more, are an essential ingredient in every
data pipeline too. To be clear, the effect of these lookups can be critical on the
outcome of a pipeline. A sudden change of address or database revision might alter
the course of pipeline artifacts radically. So it is very much in the interests of
forensic traceability to incorporate knowledge of these lookups into a pipeline process.
However, usually these lookups take place within user code---invisible and opaque.
A solution to this issue is to include them as {\em implicit connections} in a pipeline description.

What forensic details do we want? We begin with the following list:
\begin{itemize}
\item which changes triggered the recomputation?
\item Which versions were involved in recomputation?
\item If data were read from a mutable external source, say DNS, cache the response for forensic traceability.
\end{itemize}

\subsection{Data arrival policy}

A pipeline is a collection of tasks, some of which collect data, some
of which emit data, and some of which process data. Data are passed
along `links', which sit logically between tasks. Along the way, some
tasks may employ services that are formally out of band of the pipeline.

From a Promise Theory perspective, no data can be imposed on such a
design.  Data are intentionally sampled by the edge nodes (even files
dropped into an in-tray, pulled from a database or made available for
sampling by a sensor).  These samples can be promised to the next
task in some intermediate format.

\begin{itemize}
\item The usual format will be a dumb queue of values (First Come First Served).

\item Another common format is an intermediate database case, where
  data get dropped off into a reservoir, and can be tapped or
  resampled by the next stage of the pipeline, blurring the
  distinction between pipeline and client-service. This enables flows
  with different kinds of semantics, but should not be abused (e.g. to
  generate huge `landfills' of entropy that cost enormous amounts of
  processing to, such as datalakes). Timeseries databases and tools
  like Kafka can avoid loss of crucial causal ordering information.
\end{itemize}
Data inserted into such intermediate databases can be sampled by the
link agents which connect tasks, by making them `smart', and the
pipeline manager remains responsible for scheduling work. When no work
is arriving, resources can be scaled down to zero as long as cache is
not lost.

An obvious parallel is the apparent different between
publish-subscribe (pull-based processing) and imposition queues (push
based processing) such as is used for notifications.  The
proliferation of mobile devices has more or less settled the debate
about the virtues of push and pull.

\subsection{Publish-subscribe for distributed scaling}

In a data pipeline, whether we call it pushed, sampled, streamed, or batch-oriented,
data are passed in batches of some characteristic size to each process
in sequence. The connected graph of tasks forms a sparse square matrix
$D_{ab}$, where $a,b$ run over the different numbered task nodes.  The
size of data entering at each stage $a$ and wire link leading to $\ell$, $D_{\ell
  a}$, is usually different from the data leaving along different
wires leading to the agents $D_{a \ell'}$.  

In order to scale data transfers of small or large size, over long
distances, and with high latencies, we look to a publish-subscribe
(pull) model for data handovers, with a separate side channel for
instant messaging.  The advantage of this approach is that
notification messages can be pushed into a queue with predictable
consequences, and without unnecessary replication of data in the case
where the same data need to be passed to different forward branches.

\begin{figure*}[ht]
\begin{center}
\includegraphics[width=13cm]{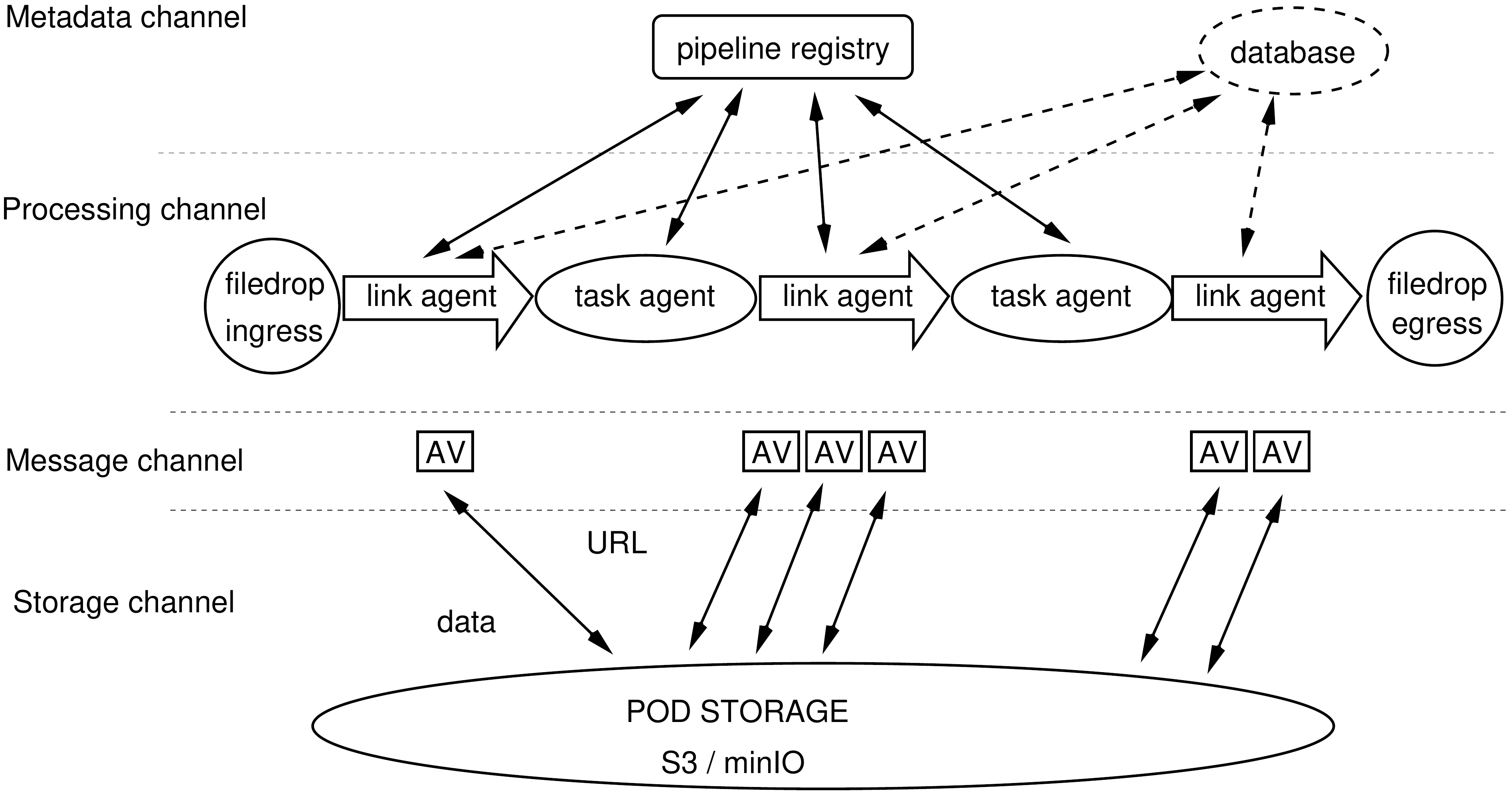}
\caption{\small Architectural elements  of Koalja, showing the three main kind of
agent: tasks, links, and the pipeline registry.\label{koaljaarch}}
\end{center}
\end{figure*}

The storage problem is particularly difficult (see next section), because one has a
priori little idea of the underlying physical constraints of the cloud
infrastructure.  The capacity of the physical network may limit or
enable the transport of data in bulk. The critical ratio: \beq \rho =
\frac{\text{Average latency of internal storage}} {\text{Average
    latency of network storage}} \eeq determines whether it is more
rational to rely on local storage copies or to load data from a remote
service. This is utterly dependent on the infrastructure choices
underlying a particular local region of cloud processing---information
which is not available to a platform to make an informed decision. We
therefore have to bet on contingencies.  Given the improvements in
network speeds, versus increases in on-board latency, we choose to
place our money on the network attached storage, assuming that cloud
infrastructure will manage this as a private channel (independent of
the channels used for web traffic and interprocess communication).

This simplifies the decision about where to store data within a
homogeneous region of cloud infrastructure, like a datacentre, but it
still leaves open what to do about regional network deficiencies.
Again, we opt to bet on the pace of network infrastructure
improvements, combined with a principle of lazy evaluation---to never
transport data that don't need to be transported.  Some may question
this, as it is commonly interpreted to mean that it is both free and
easy to push all data from into the major cloud datacentres. This is
not our conclusion.  A device at the edge of the network may produce
Terabytes of data in a few hours, most of which are junk and thus have
no business travelling outside the immediate local context in which
they were produced.

The natural choices for a commoditized approach are to limit wide area
contention by storing data as close to their source (and the context
in which they are relevant) as possible. Summarization, statistical
analysis, compression, and contextualized trending at the edge, can be
used to reduce the dimension of data prior to centralization, where
necessary.  

Apart from nodes at the user-facing edge of the cloud (where first
input, file drop and sampling take place), each smart task will sample
data from its own inputs, and passed them along the smart link that
connects it. Let's say there are inputs from three different
sources---a tuple (a,b,c). A tuple is not actual data, but URI
references to data, and other metadata.

We want to know when the task should be able to see a data tuple. The
task will try to read from the pseudo-stream, with a `get' operation.
But it wants to know if there is anything new on the channels. If
there are new data, they can be tested. If there is an arrival rate
mismatch between the different channels, the tuples will change a lot
in some values, but not others. The users can choose what smart link /
task behaviour is desired. E.g. if either b or c changes, then trigger
a new tuple in the stream of input events---otherwise don't. In other
words, changes to a do not lead to a new event (only a change in the
value of a reported).

\begin{principle}[Separation of channels by timescale]
A separate message notification channel for data arrivals may be used
for updates that are slow in arrival time compared to
the service time. 
\end{principle}
The timescale-aware approach to processing makes notifications useful
when data arrive irregularly with intervals longer than the time
for infrastructure changes. This avoids inefficient sampling of queues
that are inactive. Conversely, messaging is an overhead when arrivals
are frequent, on a shorter timescale than the time for infrastructure
changes.

\begin{principle}[Local caching close to dependents]
  Data that are chosen to be passed down the line to the next
  dependent task, will be cached local to the dependent task, for a
  policy determined length of time, if the intermediate result is
  combined with others. This facilitates later recomputation in the
  case when partial re-computations may be desirable.
\end{principle}
There is less reason to cache intermediate results that are not combined
with other data, but we have to remember that the data are not the only
`arrivals' that can trigger recomputation. A change of software version
may necessitate the recomputation of a result because it was wrong.
In the case of big data batches, the cost of recomputation along
the entire pipeline may be significant and avoidable with a 
decent caching strategy. A suitable default behaviour could be to cache
everything, but to purge the caches at different rates depending on the
risk of recomputation, for instance. This is a matter for policy.

\subsection{Storage, near and far}

Intermediate data, passed along between task stages in a pipeline need
to be stored in an expedient location under the control of the
pipeline manager.  Inside a datacentre, data and notifications about
data are normally separated into separate network channels (a storage
network and a data network, with independent capacity). This allows
data to be passed between containers `out of band' of the application,
without contending with other applications, by employing dual channels
for storage. Kubernetes plays a role here in scheduling related tasks
in local rackspace, so that traffic sharing routes are as isolated as
possible (e.g. assuming a Clos architecture). When pipelines are
federated across WAN connections, data transfers will share the
network capacity of the application, and thus compete.  Our design
principle is to regard the cost of messaging (by Annotated Value) to
be negligible.

Even in a datacentre, storage attachment within a containerized cloud
is not a trivial problem. Host machines have their own storage,
connected by an interior processor bus. In addition, they have the
possibility of mounting or accessing network storage over a typically
faster fibre channel network.  As kernel semantics are handed down
through layer upon layer of virtualization, the latency for mounting
filesystems quickly becomes longer than the expected runtime of the
container itself, which suggests the use of object storage (S3, MinIO,
etc).  To combat this, one can try to mount storage on the more
persistent shared hosts\footnote{In Kubernetes this may be handled as
  a daemonset to ensure that helper components are running on all
  machines.}, and have some kind of manager on top.  An alternative to
the use of a filesystem, with its intrusive kernel modules is to use
network object storage, such as Amazon's S3, and its open source
clones (MinIO\cite{minio} etc). We have tried both approaches during
testing.

The quantities of data involved in some processes (`big data') mean
that one wishes to avoid transporting and buffering data bundles as
far as possible.  Given the expected proliferation of the cloud, all
the way out to the `edge' of user space, it's often more expedient to
move data processing tasks than to move data. Or rather, while the
bulk of processing capacity is currently only accessible in the
centralized datacentres of the cloud, we expect this situation to be
completely rewritten in the future---there is a wealth of processing
capability that's mostly idle waiting to be harnessed at the user
edge.

When data are passed to containers for execution, they need to be packaged in a
size that can fit into local RAM (or at least local storage), and
therefore each active set for a container execution could also be cached on
local media, while larger source material may be kept more permanently
on object storage, and pointed to by a suitable URI for access on demand.

This `smart' transport avoidance can't be entirely handled by the
platform---at least, not until significant knowledge about locations
and connectivity is available to the Koalja platform. For now, users
designing their processes also need to design them to avoid
transportation of unnecessary data\footnote{MB: when moving apartment
  it requires some discipline to throw away junk before moving. Too
  often, one carries all junk to a new apartment and then throws it
  away because I can't find a purpose for it at the new location. The
  lesson is that information is contextual. If you change context, you
  need to think about its relevance to avoid being taxed on it.}.
However, in lieu of some automated decision-making, we can encourage
the proper use of scaling principles to reduce costs, by making use of storage
at the edge of the network. To avoid a conflict of
interest, for cloud providers, we expect cloud providers will simply
buy edge processing in homes and workplaces and run it as part of an
Extended Cloud. Then there is no excuse for not optimizing. This is
not just `nice to have', it is rapidly becoming a global
sustainability imperative.

\subsection{User experience}

The cloud is still much too hard for users. If a vision of utility
computing is being able to turn on a tap, or ask a home assistant to
dim the lights, then cloud computing is still at the level of having to
build a bridge across a chasm to fetch water from the local well.
Clearly much has been done to rationalize waste and simplify development,
but little has been done to make cloud easy for business purpose.

To facilitate experimentation, we have been inspired by the breadboard
model of electronics. A breadboard is a partially pre-wired patchboard
into which components can be plugged, without soldering or permanent
connection, because the connecting wires of components have
standardized profiles. It enables very quick prototyping for trial and
error, and results ultimately is a working layout that can form the
basis for a more permanent solution. The analogue of a breadboard in a
data processing environment can be enabled with basic connectivity
services and a standardized plugin mechanism.  Software containers act
as an initial pluggable entity with standard size, and the
connectivity layer can be supplied by Kubernetes, if one can solve
the latent problems of storage and networking.
One of the biggest headaches in cloud computing is the antiquated need
to specific TCP/IP parameters, especially port numbers and maps
between interior managed sub-services and the exterior promised services
that form the r'aison d'\^etre. Cloud platforms throw all of this complexity
at users, making cloud services inaccessible to swathes of users who have
no time or inclination to deal with such matters. These are all matters to
dissolve into transparency.

Ignoring all aesthetic aspects of the user experience, which one would
expect to tailor to local circumstances, we shall only focus on the task of
simplifying the data fed to the infrastructure in order to
define a process.
One version of a description, that looks from the god's eye view
of the topology is shown in figure \ref{wiring}.
\begin{figure}[ht]
\footnotesize
\begin{tabular}{rcl}
[tfmodel]&&\\
                    (in)& {\bf learn-tf}& (model)\\
                 (model)& {\bf server}&   (lookup implicit)\\
                    (in[10/2])& {\bf convert}& (json)\\
(json, lookup implicit) &{\bf predict}& (result)\\
\end{tabular}
\normalsize
\caption{A basic input language for describing process wiring.\label{wiring}}
\end{figure}
An alternative, from a local agent perspective would look more like a Makefile,
in which each agent lists the required dependencies it has been promised.
The connection between these viewpoints is straightforward. Each matching
promise of an output (+) is matched by the promise to consume it (-)
on the end of a smart link\cite{promisebook}.

Once we have a wiring diagram correctly connected using smart tasks/links, it
can be turned into a Kubernetes implementation by Koalja. This may lead to a
number of implementation decisions, some of which can be determined by
policy settings.

\begin{figure*}[ht]
\begin{center}
\includegraphics[width=11cm]{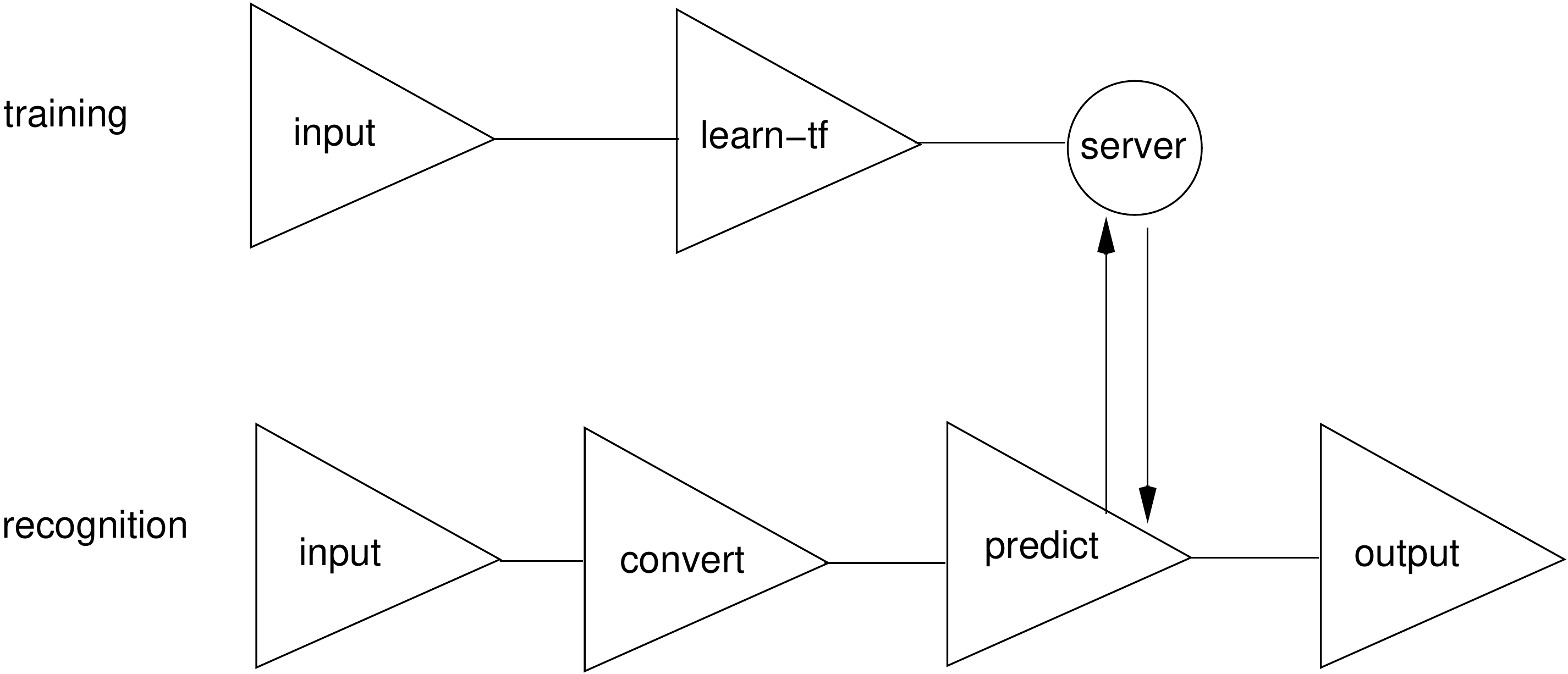}
\caption{\small 4 An example pair of pipelines, represented as a
  single data circuit.  The upper pipeline shows a training process
  for a Tensor Flow neural network, which is deployed as a service
  consulted by the lower pipeline. The lower pipeline receives sample
  images to be recognized and classified according the machine
  learning model trained by the upper pipeline.  The implicit link
  between the two pipelines is shown by the double arrow client-server
  interaction. Clearly, the timescales of the upper and lower
  pipelines are unrelated, the upper one being presumably much longer than
  the lower for stability of classification.\label{tfmodel}}
\end{center}
\end{figure*}

\subsection{`Smart' Task Agents}

Data processing requires computers to be scheduled and software to run
the logic.  Kubernetes schedules tasks as pods of containers that can
be elastically scaled. In Koalja, these become part of a `smart task'
service.  It makes sense to wrap container execution in some basic
policy-guided reasoning to avoid needless repetition of code, which is
quite difficult for non-expert users to reproduce.  One interpretation
of this is as a more general realization of the concept of Network
Function Virtualization (NFV) or Service Chaining promoted by
networking bodies as a model of cloud services.

\begin{figure}[ht]
\begin{center}
\includegraphics[width=7.5cm]{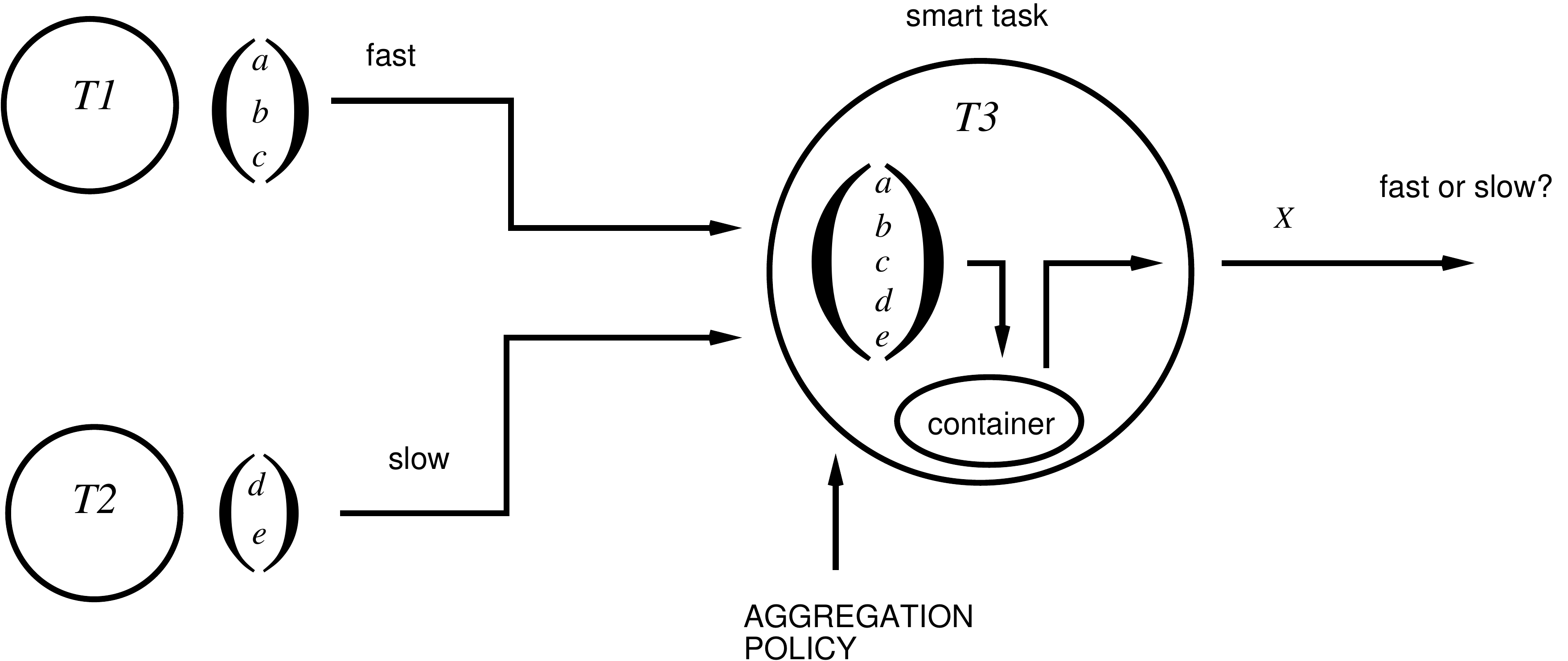}
\caption{\small Aggregation of data from multiple input channels, e.g.
  collecting data from multiple weather sensors for a complete sample
  set.  Some sensors (e.g. wind speed) may take longer to arrive than
  others (e.g. temperature). Should the pipeline wait for all the data,
several repeated measurements (as an time-series of values or as a sliding window).
There are several common possibilities for coordinating and composing data, relative
to the assumptions of user code.\label{aggregate}}
\end{center}
\end{figure}

Not all data processing proceeds as a simple bucket chain.
Manufacturing processes assemble parts from several supply chains, and
the same is true on data processing.  Similarly, some parts of a
process involve calling upon services on demand, such as help from an
operator, look-up of addresses, consulting with specialized
recognition services which are trained by their own independent
pipelines, weakly coupled.  Smart tasks therefore arrange for data to
arrive at user containers as sets of `Annotated Values' (an internal
representation of a single execution set, yet to be converted into a
consumable form). The annotation refers to metadata used in tracking the artifact.
The value is in fact a message that points to a storage location for the data,
thus avoiding the need to send actual data through from link to link as a queue.
The annotations include
\begin{itemize}
\item A unique identifier for forensic tracing.
\item The source task that produced it as output.
\item Pointers to the links and storage locations of the actual data.
\item A local timestamp for the creation, which refers to the clock of the source agent.
\end{itemize}
Annotated values may arrive from more than one
incoming smart link, in which case the payloads can be aggregated into
tuples in a number of ways (see figure \ref{aggregate}).  These
correspond to blocking and non-blocking policies, sliding windows,
filling criteria, and so on, as we discuss below.  The task agent's
common wrapper services thus promise to assemble {\em snapshots} or
policy-determined data sets as locally stored file chunks that can be
fed to a container execution command in the form:
\begin{alltt}
<USER CODE> <ARGV list>
\end{alltt}
A user needn't even know how to build a software container for to wrap
his or her code. That process is also automatable.
A few common cases for how data are aggregated can be handled by a smart
platform. The task agent has the responsibility to  wait for data from
its incoming links and assemble execution sets of annotated values to construct
the arguments for a single execution (see figure \ref{aggregate}).

We can assume that all data are passed as a stream of statically or
dynamically typed data tuples (see figure \ref{aggregate}), delivered
as a stream of Annotated Values per incoming link, represented in the
diagrams as parenthesized column vectors `(..)'. Thus every data
stream is a uniformly populated sequence of input vectors.  A smart
link offers $(a,b,c,d…)$ and the receiver smart task may select any
sublanguage $(b,d,....)$ at its discretion---this is similar to the
interactions used in GraphQL\cite{graphql}. This selection can be
performed is best made the user application. Koalja simply merges the
stream of possible values.

A link agent (as described in the next section) then delivers each
Annotated Value (AV) as a labelled text file, available to a task as a
local file, handing it over by name as a file to argv.  Since tasks
are assumed to be wrapped in containers, there is no other natural communications channel for
passing values.  The source and destination for AVs may be files,
or reads and writes to a database or message queue. A task specification 
can have a very simple form: (input wires, container, output wires):

\begin{alltt}
\small
 (input1 input2 ....) 
   taskname 
      (output1, output2 ...)
\end{alltt}
Another level of policy can be used to provide some common services
to users. The inputs represent tuples of files that are collected as a
`snapshot' set, according to a policy.  A named input may have a
buffer size, representing the minimum number of AVs required to
execute the container, e.g.
\begin{alltt}
\small
 (input1[5] input2 ....) 
    taskname 
      (output1, output2 ...)
\end{alltt}
Koalja can then enable services like sliding windows to be managed by the 
task service itself, rather than burdening user code with these common issues.
A snapshot is thus a set of input files to be substituted for argv in the task container.
Even for a single input, there are two ways of aggregating data: 
\begin{itemize}
\item Each input (or input buffer) reads new incoming data, using them only once to produce an output. 

\item An input maintains a sliding window in which a subset of values are replaced, like a queue.
\end{itemize}
Sliding windows and buffers can have different sizes for different incoming data, e.g.
we might need ten stream data to compute a running average, but only a single value from
another process to scale them by.
When several inputs are combined, they form a tuple of inputs, each of which might be a buffer:

\begin{alltt}
\small
 (input1[10] input2[1] ....) 
   taskname 
     (output1, output2 ...)
\end{alltt}

A policy for constructing snapshots determined
how to advance the AVs, arriving on each of the links, to form the tuple execution sets.
There are two main cases that can occur since each link may bring new data at different
rates (see figure \ref{aggregate}):
\begin{itemize}
\item Some of the inputs have new data, but we want to recompute
  immediately; e.g. when compiling software, changes to only a few
  files are common, but all files are needed to built the output
  software. So then, older values of some inputs are combined with new
  values for changed files. The result is a mixed policy (partial swap
  of new for old).

\item Wait for a sufficient number of new values to arrive on all
  incoming links before recomputing an output.  This has several
  variations (min/max or exact required number), with different
  consequences for what the task code will be fed.
\end{itemize}
Snapshot policy may also promise a rate control to avoid needless unintended recomputation,
and the possibility of Denial of Service attacks on the inputs.
In practice, only a few of these policies are likely to be common, but they
can be straightforwardly provided.
We refer to the policy choices by the internal names:
\begin{itemize}

\item {\em All new}, means that there is no reuse of values in a snapshot.
Each snapshot is formed from a non-overlapping set of completely fresh data.
This is what usually happens in a stream.
The type of the aggregation is a direct vector sum of the types belonging to the
incoming links.

\item {\em Swap new for old}, means that if new values appear on a link, fresh
values will be assembled into a snapshot, but where there are no new values, previous
values will be used. This is like the aggregations in a Makefile.
The type of the aggregation is a direct vector sum of the types belonging to the
incoming links.

\item {\em Merge}, means that data from multiple links will be aggregated in a First
Come First Served order into a single scalar stream. For this to happen, the data values
must be of the same type.

\end{itemize}
In addition to these selections, there is the possibility to build 
`sliding windows' over the last $N$ value sets. When a window slides by
two positions, for instance, with a buffer of $N$ values, two new
values are read (according to one of the policies above) and the two
oldest values fall off the end of the snapshot set, ensuring a
constant number with two refreshed values, e.g. \verb+input[10/2]+
for a buffer of 10 values, sliding 2 positions at a time.  This is useful for
computing moving averages etc.

\subsection{`Smart' Link Agents}

Smart links marshal the data as files for the task code.  The logical
connection between the outputs from one task and the inputs of the
next are handled by these link agents.  The physical storage of
intermediate results, which are assumed to be temporary in nature, but
can be cached for re-use according to policy, is object storage---as discussed above.
When triggers for recomputation of a pipeline occur due to
changes out of band of the dataflow, process identifiers may need to
go back to earlier values and recompute them. Changes include:
\begin{itemize}
\item Incoming sample updates.
\item Software Updates.
\item Service dependency updates.
\end{itemize}
For the latter two cases, a change may be due to software errors,
indicating that recomputation is needed.  Smart links can simply
behave as if one can `roll back' the feed. This may depend on
there being no new arrivals to handle in real time; on the other hand,
we should remember that, as a cloud process, several computations can
be handled in parallel as long as sufficient infrastructure resources
are provided.  `Big data' require `big' processing power and a lot of
energy to generate. Storing results is thus most likely far cheaper
than regeneration of results in cases where data results need to be
re-used.

For example, in the processing of build pipelines, using `make' or
some other tool, it's unnecessary to recompile binaries that are
unchanged in order to link them with updated files. Sparse updates
allow enormous savings in data processing, as tools like Make has
exploited for decades.

If we apply this thinking to the twin pipeline model in figure
\ref{tfmodel}, we see all the variations of updating around the
process. Training of the data sets from a pool of long-term
accumulated data is pulled from accumulated sources on demand (like a
recursive make build). The arrival of data (say faces) to be
recognized by the model is a random arrival process (a stream) which
is much faster and in a forward direction.  Then the software
algorithms in the running containers may also be updated, which may
trigger the recomputation of certain matches (perhaps faces that could
not be recognized on the first time around, or were missed due to bad
formatting). The user clearly plays a role in this process, deciding
when policy based updates to software and data should be changed,
and deciding which data to check again after a software update.
\begin{figure*}[ht]
\begin{center}
\includegraphics[width=12cm]{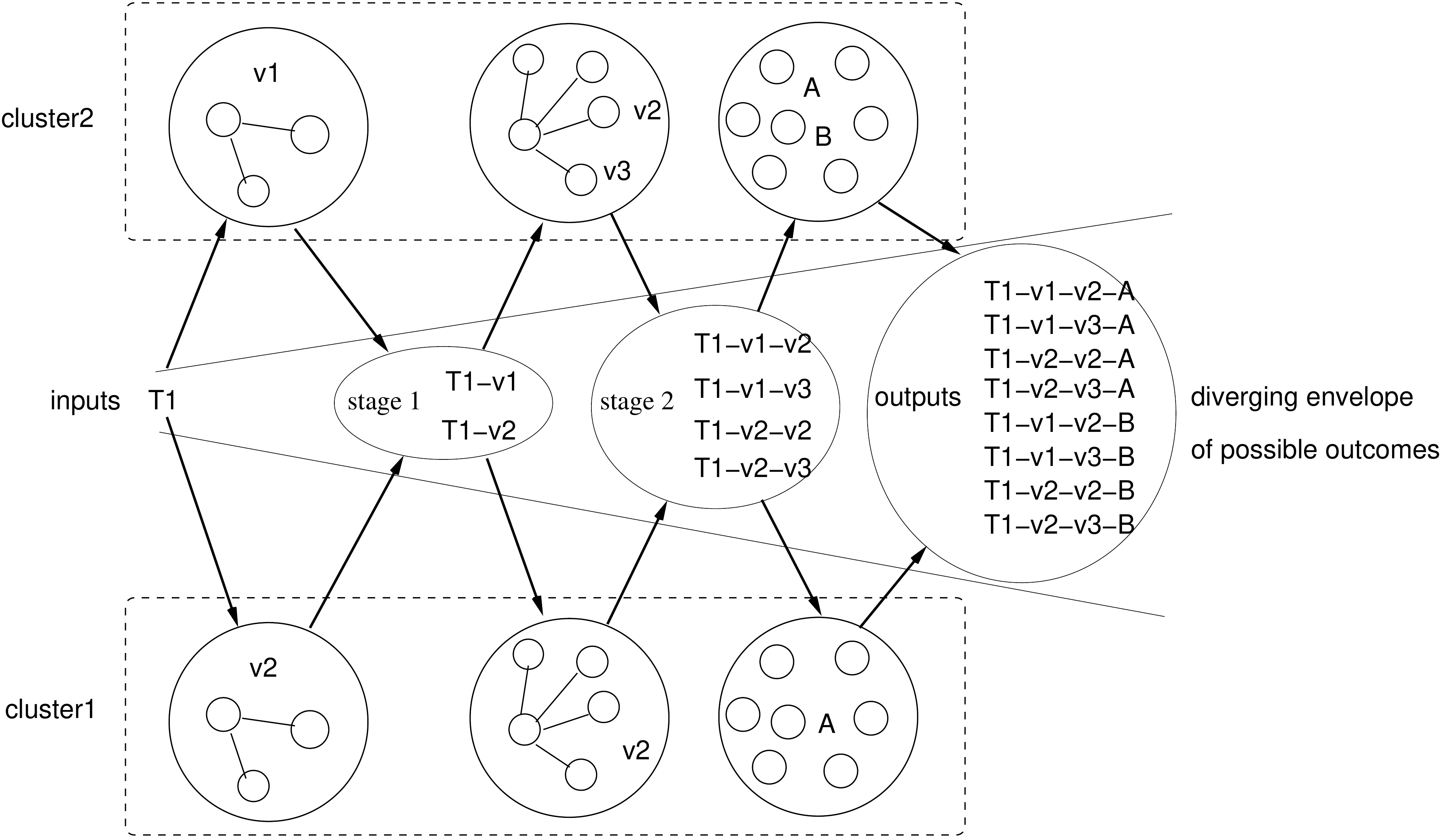}
\caption{\small As data are shifted around processes in 
possibly parallel pipelines, formed from elastically scaled
agents running software, which is changing in real time, 
we need to be able to see the causal travel documents
of the data to know exactly what led to outcomes.\label{pipeline2}}
\end{center}
\end{figure*}

\subsection{Continuous improvement and wireframing}

Push pipeline models are a classic `throw it over the wall' model of
data processing: downstream stages are triggered by upstream
arrivals---by imposition. Resources downstream have no control over
their expected load or resource requirements in such a model. Although
one could argue this is a natural side-effect of delegation, it can
lead to catastrophic failures.  When data processing is ordered by
selecting a desired end state (like a Makefile, a CFEngine
configuration, or a Kubernetes deployment), the problem is reversed
because one has precise knowledge of what resources are required.

A related issue turns out to be how one handles testing and continuous
improvement processes.  Semantic testing of a data processing pipeline
needs data, but using real data may be wasteful, where full resolution
leads to large and expensive resource burdens.  The way around this is
to develop code on small amounts of data and iterate by trial and
error.  The dilemma is that such small data may have to be generated
artificially, and suffer from test biases, wasting time and leading to
mistakes. 

In the spirit of breadboarding, we expect that users with need or want
to experiment before building a robust business process. It's in the
nature of cumulative processes that getting the first rough working
parts together is the hardest stage, then tuning its details can be a
pleasurable experience. So the approach we take should allow simple
cumulative successes.  A pipeline platform can help here.  We can
apply quality assessment and iteration with modifications to user code
within an interactive loop.  Continuous delivery systems need to be
able to distinguish between artifacts however---here the tooling can
help enormously to trace the precise versions and timings that led to
a particular outcome.

When data sizes are large, we might waste a lot of time recomputing
values to iterate over improvements to the pipeline user code.  There
is plenty of scope for caching intermediate results---as `make'
processes do.  However, its useful to trace data all way through to
test handovers too.

The scenario is analogous to the rendering of computer graphics for
games or film. While the details of character and storyline are being
ironed out, it's useful to work with wireframe models.  Sub-sampled
outlines of the data may be sufficient for debugging.  This concept
has not previously been applied to data pipelines at the platform
level to our knowledge, but it seems more natural than managing a
separate, out of band process, to generate artificial and potentially
biases data sets for testing. The only real tests are real data.

\begin{figure*}[ht]
\begin{center}
\begin{alltt}
\tiny
New process timeline for ( myApp_name21.2.3 ) originally started as pid  17778 

Unix clock context              | root --> NOW,delta  Comment indented by subtime
------------------------------------------------------------------------------------------
2019-06-03 13:40:04 +0200 CEST  |    0 -->   1,1      MainLoop start 
2019-06-03 13:40:04 +0200 CEST  |       ->   1,2        [function: main] 
2019-06-03 13:40:04 +0200 CEST  |    1 -->   2,1      Beginning of test code 
2019-06-03 13:40:04 +0200 CEST  |       ->   2,2        [remarked: : Start process] 
2019-06-03 13:40:04 +0200 CEST  |       ->   2,3          [go package: cellibrium] 
2019-06-03 13:40:04 +0200 CEST  |       ->   2,4            [btw: example code] 
2019-06-03 13:40:04 +0200 CEST  |       ->   2,5              [remarked: : look up a name] 
2019-06-03 13:40:04 +0200 CEST  |    2 -->   3,1      code signpost X 
2019-06-03 13:40:04 +0200 CEST  |       ->   3,2        [intent: : open file X] 
2019-06-03 13:40:04 +0200 CEST  |       ->   3,3          [file: /etc/passed] 
2019-06-03 13:40:04 +0200 CEST  |       ->   3,4            [dns lookup: 123.456.789.123] 
2019-06-03 13:40:04 +0200 CEST  |       ->   3,5              [btw: xxx] 
2019-06-03 13:40:04 +0200 CEST  |       ->   3,6                [coroutine: main] 
2019-06-03 13:40:04 +0200 CEST  |    3 -->   4,1      Run ps command 
2019-06-03 13:40:04 +0200 CEST  |    3 go>   5,1      TEST1--------- 
2019-06-03 13:40:04 +0200 CEST  |       ->   4,2        [btw: /bin/ps -eo user,pcpu,pmem,vsz,stime,etime,time,args] 
2019-06-03 13:40:04 +0200 CEST  |       ->   5,2        [btw: Testing suite 1] 
2019-06-03 13:40:04 +0200 CEST  |       ->   5,3          [intent: : read whole file of data] 
2019-06-03 13:40:04 +0200 CEST  |       ->   5,4            [file: file://URI] 
2019-06-03 13:40:04 +0200 CEST  |       ->   5,3          [remarked: : file read failed] 
2019-06-03 13:40:04 +0200 CEST  |       ->   5,4            [system error message: open file://URI: no such file or directory] 
2019-06-03 13:40:04 +0200 CEST  |       ->   4,3          [remarked: : Finished ps command] 
2019-06-03 13:40:04 +0200 CEST  |    3 go>   6,1      Commence testing 
2019-06-03 13:40:04 +0200 CEST  |       ->   6,2        [remarked: : Possibly anomalous CPU spike for this virtual CPU] 
2019-06-03 13:40:04 +0200 CEST  |       ->   6,3          [anomalous CPU spike: CPU 22117.000000 > average 22115.000000] 
2019-06-03 13:40:04 +0200 CEST  |    6 -->   7,1      A sideline to test some raw concept mapping 
2019-06-03 13:40:04 +0200 CEST  |       ->   7,2        [btw: Commence testing] 
2019-06-03 13:40:04 +0200 CEST  |    7 -->   8,1      End of sideline concept test 
2019-06-03 13:40:04 +0200 CEST  |       ->   8,2        [btw: Commence testing] 
2019-06-03 13:40:04 +0200 CEST  |    6 go>   9,1      Starting Kubernetes deployment 
2019-06-03 13:40:04 +0200 CEST  |       ->   9,2        [btw: Commence testing] 
2019-06-03 13:40:07 +0200 CEST  |       ->   9,3          [remarked: : Starting kubernetes pod] 
2019-06-03 13:40:10 +0200 CEST  |       ->   9,3          [remarked: : File drop in pipeline] 
2019-06-03 13:40:13 +0200 CEST  |       ->   9,3          [remarked: : Querying data model] 
2019-06-03 13:40:16 +0200 CEST  |       ->   9,3          [remarked: : Submit transformation result] 
2019-06-03 13:40:19 +0200 CEST  |    6 go>  10,1      The end! 
2019-06-03 13:40:19 +0200 CEST  |   10 -->  11,1      Show the signposts 
\end{alltt}
\caption{\small A Local checkpoint log, with interleaving and branching timelines, as discussed
in \cite{observability}.\label{map1}}
\end{center}
\end{figure*}

\begin{figure*}[ht]
\begin{center}
\begin{alltt}
\footnotesize
<begin NON-LOCAL CAUSE>
(program start) --b(precedes)--> "MainLoop start"
  (MainLoop start) --b(precedes)--> "Beginning of test code"
     (Beginning of test code) --b(precedes)--> "code signpost X"
        (code signpost X) --b(precedes)--> "Run ps command"
        (code signpost X) --b(precedes)--> "TEST1---------"
        (code signpost X) --b(precedes)--> "Commence testing"
     (Commence testing) --b(precedes)--> "The end!"
        (Commence testing) --b(precedes)--> "[remarked: : Possibly anomalous CPU spike for this CPU]"
        (Commence testing) --b(precedes)--> "A sideline to test some raw concept mapping"
        (Commence testing) --b(precedes)--> "Starting Kubernetes deployment"
        (Commence testing) --b(precedes)--> "The end!"
        (Commence testing) --b(precedes)--> "[remarked: : Possibly anomalous CPU spike for this CPU]"
        (Commence testing) --b(precedes)--> "A sideline to test some raw concept mapping"
        (Commence testing) --b(precedes)--> "Starting Kubernetes deployment"
     (The end!) --b(precedes)--> "Show the signposts"
        (A sideline to test some raw concept mapping) --b(precedes)--> "End of sideline concept test"
     (The end!) --b(precedes)--> "Show the signposts"
        (A sideline to test some raw concept mapping) --b(precedes)--> "End of sideline concept test"
     (code signpost X) --b(may determine)--> "[dns lookup: 123.456.789.123]"
        (TEST1---------) --b(may determine)--> "[file: file://URI]"
        (TEST1---------) --b(may determine)--> "[file: file://URI]"
<end NON-LOCAL CAUSE>
\end{alltt}
\caption{\small An invariant concept map of an instrumented data pipeline that may
include user code, as described in \cite{observability}.\label{map2}}
\end{center}
\end{figure*}

\subsection{Metadata collection}

One of the issues that businesses and organizations struggle with, in
a globalized Internet, is how to track data between applications, when
there is no common model for sharing data. Issues of data provenance
and accessibility outside of specialized security zones are quite hard
to handle. This is not just a matter of encrypting data end to end, but
managing the access controls of intermediate storage, caches, and so forth.

As data move between tasks, through storage and caches, they trace out
a path history of interactions that may capture the causal history of
the outcome. This path of interactions leads to the three kinds of
story mentioned in section \ref{meta}. Checkpoint logs are handy for
debugging and for forensic reconstruction from a developer viewpoint.
The map of data invariants is handy from a total model perspective,
which applies both to developers and end users who consume data from
pipelines.  Perhaps most important, however, is the traveller's
journal: every data packet's travel documents get stamped according to
the journey taken. These documents are associated with a unique ID,
and the documents are kept in a secure location by the pipeline
manager.  As data move, metadata of the path history is accumulated
and grows in this pipeline manager's registry. In a long pipeline, the
possibility for combinatoric outcomes is quite large (see figure
\ref{pipeline2}), which indicates the virtual impossibility that users
would be able to guess these combinations during debugging or forensic
analysis later. This doesn't mean that the metadata need to be large,
but it does mean that the amount of data overhead kept versus the
number of possible reconstructions becomes tiny. In other words, it
is cheap to keep traveller log metadata for every packet, compared to
the expense of trying to reconstruct by inference at a later date (cf:
the mashed potato theorem in \cite{observability}).  It's therefore
highly expedient to collect such metadata for later forensic analysis,
debugging, or certification---especially where there may be
constraints on the allowed paths for data to follow, e.g.  US data
cannot leave the virtual boundary of the US.

The three kinds of story, referred to above, that we want to be able
to tell about data processing come to life in a simple set of data
collection semantics, described in \cite{observability}. A library of
access points for users to equip their own code with logging
functions, to integrate into the big picture, is easily done should
they be so inclined (see \cite{observability}). One can also rely on
the smart wrappers to do the heavy lifting.

This may seem like a trivial matter, but present day logging
capabilities are extremely primitive and poorly suited to debugging.
They are also entirely developer focused. By using a custom designed
system of wrappers, designed to capture causal history in an Extended
Cloud, one can achieve considerable data compression of logs, as well
as keep them in a secure location. The benefit is that one avoids the
need to scour datalakes, based on ad hoc and ill-formed search
criteria, to glean knowledge haphazardly about what happened during
data processing. Many enterprises are concerned with liability issues
(who to blame when something goes wrong), but we prefer to think of
the metadata functions as a more preventative form of instrumentation.
If someone is monitoring the human choices, and the system `autopilot'
of all the underlying routing choices, one can debug such issues before
damage results.

In this connection, it's interesting to note a benefit of the wireframing
model discussed above. The most basic execution of a data pipeline is to
send to real data at all. By sending ghost batches through a pipeline,
we can expose where data actually end up being routed, in test runs prior
to exposing to real data (`trust, but verify').

We've discussed the technical issues around observability and tracing
in a separate paper \cite{observability}. Some example outputs of the
checkpoint log and semantic map are shown in figures \ref{map1} and
\ref{map2}, for completeness. Thanks to a strict data format, special
tools can be provided for querying these logs, so that users don't
need to rely on matching text against expensive regular expressions
and hoping for the best. We refer readers to \cite{observability} for
more details.

\section{Workspaces and IoT}

The practical uses of virtualization (wrapping services in layers of
abstraction, which can make additional promises) have proven to be
practical again and again in information systems. One natural benefit
is to control what information is transparent or opaque to users
by placing a system monitor between interior details and exterior
access. Currently, cloud computing's natural barrier is an ad hoc
one: the walls of the datacentre that contains its machinery.
Inside, there may be virtual separation between different clients
using virtual private networks, etc. As we reach the age of edge
computing, however, these approaches need to be extended so that
virtual networks can extend beyond the walls of a datacentre.

Process federation will play an increasing role in data processing,
because cloud bravado simply cannot scale to the kinds of processes we
expect just over the horizon.  For example, a modern `smart' vehicle
may produce terabytes of data on every journey, helpful for
diagnostics as well as for mapping of roads and traffic
conditions---most of which is transitory, and not worth keeping after
screening. A good fraction of the data will concern local conditions:
road defects, traffic delays, etc, which are {\em local} phenomena
whose context does not extend very far.  The data that remain, are
destined for different consumers, so there is no sense in uploading it
to a central point, only to be sent somewhere else, or even back again
(after sorting).  It is not only impractical but would utter madness to
upload such amounts from every vehicle to centralized locations every
time a vehicle reached a charging station. This is where cloud needs
to get a lot smarter.

In previous work, Burgess et al. described the concept of `smart
workspaces' in which embedded workplace services data enabled
virtualized, privatized, and commoditized sharing of data, based on a
draft architecture\cite{workspaces,ietfiot2} originally modelled on
CFEngine\cite{burgessC1}.  At the time, no suitable platform was
available to realize this vision. The arrival of Kubernetes and
container scheduling made a basic implementation plausible.
The long term story for Koalja is therefore quite interesting.

Human workspaces are a part of everyday life in the human realm We can
bring some of that familiarity into data processing too.  Delegation
and federation of process roles are a natural part of this.  A pure
pipeline abstract is too simplistic to handle this convincingly, but
if we consider the wiring or plumbing technology as an opportunity
analogous to the pipes in our homes and workplaces, then integrated
plumbing opens the door to opportunities like integrated toilets, central
heating, air conditioning, and running water.
Revealed in this light, it seems shocking that such basic services
are only just being provided in the scope of IT tools.

\begin{figure}[ht]
\begin{center}
\includegraphics[width=7.5cm]{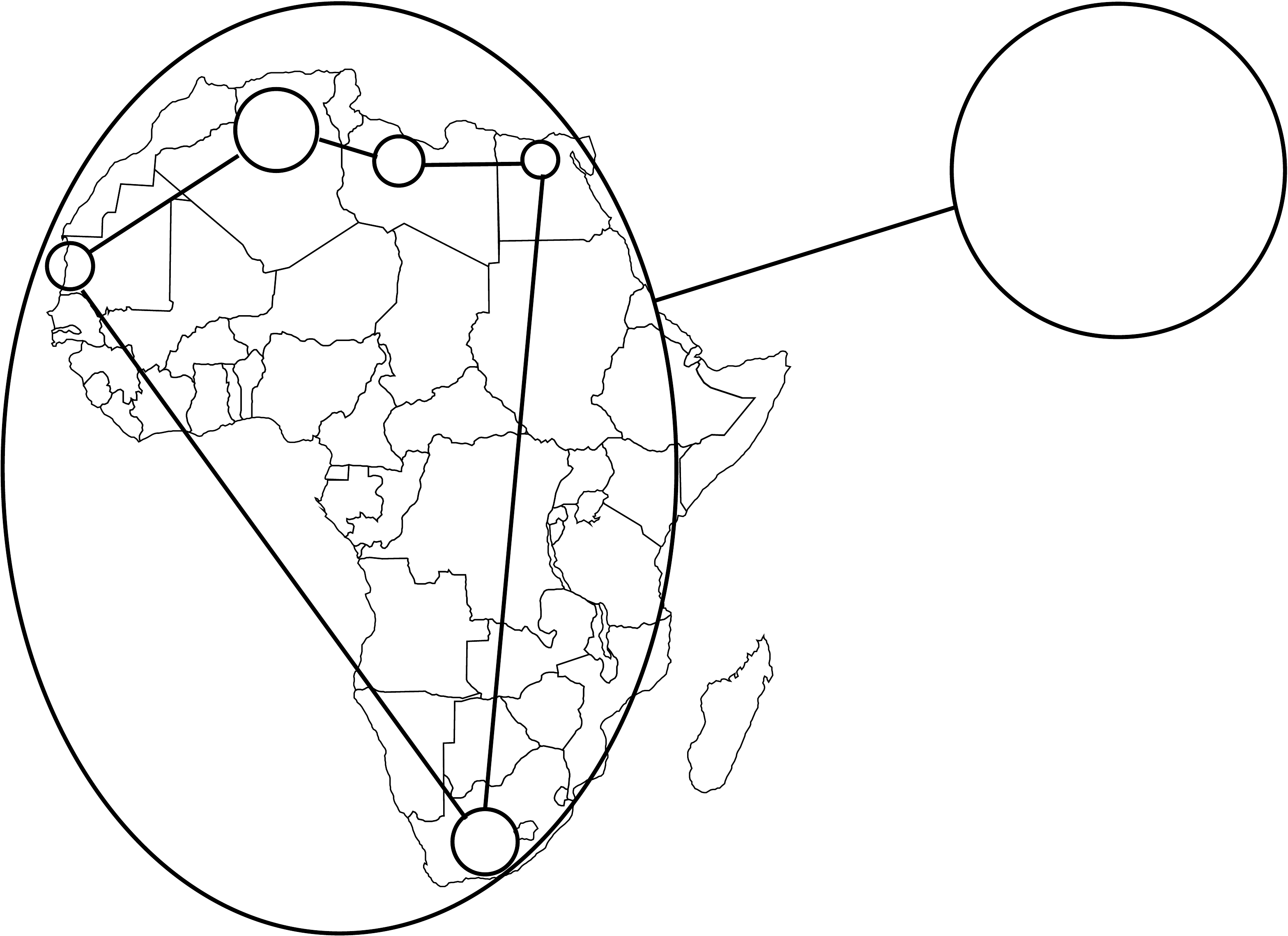}
\caption{\small The virtual access boundary for a data wiring may be
  based on many criteria, as a matter of policy. Sometimes we may want
  the same process to span different geographical regions, some even
  in motion with respect to others. Other times, we may want to separate data by function in
  the same room.\label{workspace}}
\end{center}
\end{figure}

In a globalized world, international organizations often struggle with
data sharing between sovereign realms. For example, consider a telecom
operator based in Europe, with arms in Asian and Africa. Monthly
aggregation of statistics and sales data from an African state should
never leave its country of origin, but summarized data can be
aggregated from all countries to head office (see figure
\ref{workspace}). This kind of problem is likely to become more common
as societies become more savvy about data sharing, and laws and
regulations increase in number and complexity. We need to get ahead of
the problem.

\begin{figure}[ht]
\begin{center}
\includegraphics[width=7.5cm]{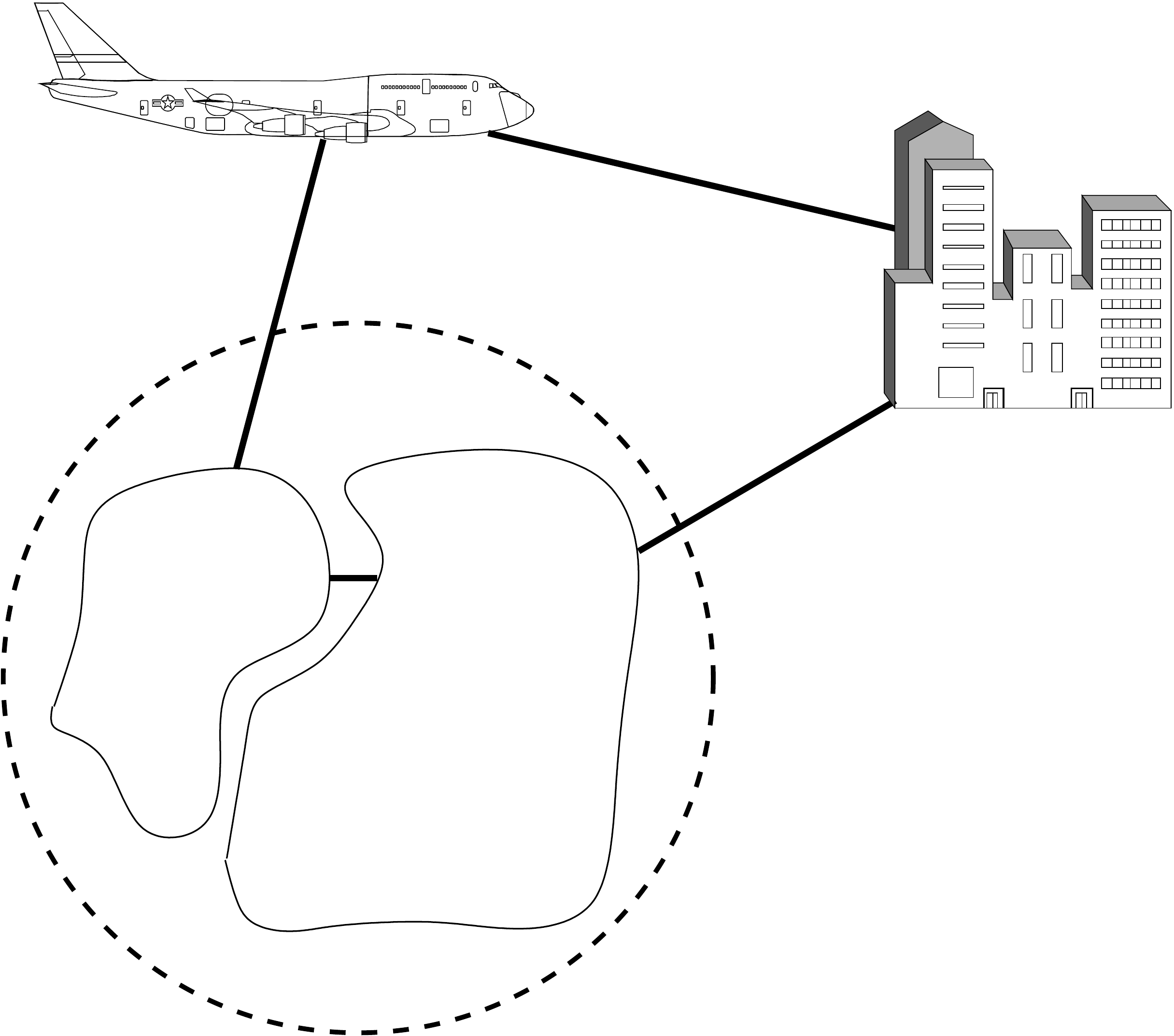}
\caption{\small The virtual access boundary for a data wiring may be
based on many criteria, as a matter of policy. Sometimes we may want
the same process to span different geographical regions. Other times
we may want to separate data by function in the same room.\label{workspace2}}
\end{center}
\end{figure}

In an abstract workspace, users would be able access shared data, but
simultaneously protect it from wider release, regardless of
geographical constraints---enabling basic federation of roles and
resources. Specialized method services associated with the data could
be embedded in an Object Oriented manner, but workspaces could also be
made to overlap as `friends', through a form of Role Based Access
Control---thus avoiding the limitations of a hierarchy of mutual
exclusion zones.  Koalja's design, which follows CFEngine's
overlapping-set-based model of inclusion, enables data plumbing to be
transformed into flexible, purpose-specific, multi-region, long-term
collaborations---even locations that don't have fixed geographical
relationships, like trains, ships, and planes (see figure
\ref{workspace2}).

The metadata subsystems we have developed, based on
Cellibrium\cite{cellibrium}, serve this function well. As sources are
more distributed and clocks smeared over multiple timezones, it
becomes important to have an interior view of timelines to understand
processes and what makes them tick, so to speak.
There is much more to say on these issues at a later date.

\section{Summary}

Our conception of data plumbing is very much like a breadboarding
model of electronics, with added benefits that automation can bring to
software. Koalja is a simple model with Smart Tasks connected by Smart Links,
instrumented with enterprise grade metadata.

Koalja feels like a natural refactoring of obvious issues for the next
phase of cloud computing where we break out of the constraints of
rack-based computing in datacentres, and emerge into the wider world,
at the so-called ubiquitous edge.  A basic level of de facto
standardization is needed to enable this.  Today, every organization
is doing data processing differently, and the problems begin during
mergers, acquisitions, and partnerships. All the burdens of compliance
with each others imposed APIs gets thrown onto end users, slowing
innovation.

Even with a platform like Koalja, there is plenty of scope for
plugin developments to simplify processes further, with third party
help. Data enabled services lead to obvious complexity, and the best
way to help with that is still to engage smart actors to make the
problems go away, allowing users to focus on what they are good at.

{\bf Acknowledgment:} we are grateful to Petar Maymounkov and Joseph Jacks
for discussion and collaboration.

\bibliographystyle{unsrt}
\bibliography{spacetime}

\end{document}